# First results from beam emission spectroscopy in SPIDER negative ion source[a]

M. Barbisan,[1] B. Zaniol[1], R. Pasqualotto[1], G. Serianni[1] and M. Ugoletti[1]

[1] *Consorzio RFX (CNR, ENEA, INFN, University of Padova, Acciaierie Venete SpA), C.so Stati Uniti 4, 35127 Padova,*

*Italy*

The SPIDER experiment, part of the Neutral Beam Test Facility (NBTF) at Consorzio RFX (Padua, Italy), is the prototype of the negative ion source for the ITER neutral beam injectors; the source is coupled to a 100 kV three-grid acceleration system. A Beam Emission Spectroscopy (BES) diagnostic was installed in SPIDER to study and optimize energy distribution, aiming, uniformity and divergence of the $H^-/D^-$ beam extracted from the source. The diagnostic is based on the analysis of the Doppler shifted $H_\alpha/D_\alpha$ light emitted in the interaction between the beam particles and the $H_2/D_2$ molecules of the background. In 2019 the BES diagnostic in SPIDER was installed and calibrated, and it allowed to characterize the first hydrogen beams extracted from the SPIDER source, in cesium free conditions. The number of active beamlets composing the beam was reduced from 1280 to 80, affecting the BES diagnostic capabilities. This paper presents the BES diagnostic setup and discusses the first collected results. Under limited extracted current density (~10 $A/m^2$) and ion energy (≤35 keV), no significant vertical beam deflection caused by the magnetic filter field in the source was detected. In some cases the beamlets were observed to be elongated in horizontal direction; beamlet divergence values down to 20 mrad and 30 mrad e-folding were measured in vertical and horizontal direction, respectively; the intensity of the Doppler shifted radiation was found to be strongly correlated to the beam current and to the beam divergence. The progressive compensation of beamlet deflections (caused by electron suppression filter fields) with increasing voltage in the extraction gap was studied.

**I. INTRODUCTION**

Among the additional heating systems of the ITER experiment there are Neutral Beam Injectors (NBIs). Each of these devices will deliver a beam of 16.7 MW, composed by H/D particles accelerated at 1 MeV [1-4]. The presently most reliable and feasible way to produce a beam of particles at this energy is to start from a $H^-/D^-$ beam, and neutralize the ions by collision with molecules of the background gas along the beam path. In ITER NBIs, the $H^-/D^-$ source and the ion acceleration system must fulfill several demanding requirements: negative ions will be extracted from the source with a current density of 350 $A/m^2$ (H)/285 $A/m^2$ (D), from a total extraction area of 0.2 $m^2$, and with a ratio of co-extracted electrons below 0.5 ($H^-$)/1 ($D^-$) [1-4]. The development of ITER NBIs is being performed in the Neutral Beam Test Facility (NBTF) of Consorzio RFX Padua. The NBTF hosts two experiments[5,6] : MITICA, i.e. the prototype of the full NBIs, presently under construction, and SPIDER, the prototype ion source coupled to a 100 kV three grids acceleration system, in operation since 2018[4] . The ion source must be able to produce a beam with current non uniformities below 10 %; in the perspective of preserving the NBI beamline components from excessive heat loads, the beam divergence shall be not greater than 7 mrad e-folding, and beam aiming shall also be monitored[1]. To minimize beam divergence, the fraction of negative ions neutralized inside the acceleration system (mainly because of collisions with gas molecules) has to be minimized, as they are not accelerated and focused by the electric fields in between the grids. Three diagnostics are available in SPIDER to study the beam properties[7]: IR imaging of the diagnostic calorimeter STRIKE[8], visible imaging of the H⁻/D⁻ beamlets extracted from the grids apertures[9] and Beam Emission Spectroscopy (BES)[7,10,11], which analyzes the spectrum of the Balmer $H_\alpha/D_\alpha$ light emitted by the beam[12,13]. The light is collected along multiple Lines Of Sight (LoSs), at an oblique angle with respect to the beam trajectory, to measure beam aiming and divergence and, with multiple LoSs, spatial profiles of beam intensity. This technique is being widely used in several negative ions sources for fusion devices. [14-17]

---

[a] Author to whom correspondence should be addressed. Electronic mail: marco.barbisan@igi.cnr.it

The initial design and simulation of the BES diagnostic for SPIDER was presented by Zaniol et al.[10]. The present paper first recalls the principle of operation of the BES diagnostic and the analysis algorithm of the collected data (sec. II); it then shows the final design and integration of the BES diagnostic in SPIDER (sec. III). Finally, the paper presents the physics results obtained by the diagnostic in the first SPIDER experimental campaign with beam extraction, which took place in 2019 (sec. IV). In this initial phase the SPIDER source was operated in pure hydrogen, without evaporating Cs to improve negative ion production.[18-20] In absence of Cs the extracted current density and acceleration voltage are reduced[22], but understanding and management of the source were simplified; in any case, in these conditions beam divergence cannot be minimized below 7 mrad e-folding as expected[1].

During the 2019 SPIDER experimental campaigns, a molybdenum mask was installed on the grid interfacing the source plasma with the acceleration system[21]. Only 80 out of the 1280 grid apertures were left open, to limit the gas outflow from the source and minimize the pressure of the hydrogen gas surrounding the source, so reducing the risk of arcs and breakdown caused by radiofrequency power input.[4] As a negative consequence, the reduction of extracted beamlets, added to the already low beam current density in the absence of Cs, forced to keep the time resolution of the diagnostic to 0.2 Hz. Nevertheless, the geometry of the active beamlets was designed so that a relevant number of them was observable by the BES lines of sight, allowing a sufficient sampling of the beam properties in horizontal and vertical direction (sec. III). The reduction of beamlets also gave benefits: the rows of beamlets are subject to alternated horizontal deflections, so if lines of sight picked up light from multiple adjacent rows, the measurements of average beamlet divergence may be overestimated; it would be never totally possible to separate beamlets deflections from beamlet divergence.[14] In the present configuration, instead, active beamlets are never close enough to cause the mentioned issue; the beamlets divergence estimations here presented are then more accurate than it would be possible in a 1280 beamlets configuration.

## II. PHYSICS BACKGROUND

A BES diagnostic consists of one or multiple collimators, collecting the $H_\alpha/D_\alpha$ light produced in the reactions between the fast beam particles ($H^-/D^-$, H/D, $H^+/D^+$) and the gas molecules, which in first approximation can be assumed at room temperature. While the photons emitted by the molecules dissociated in the collisions are received with negligible Doppler shift, the photons produced by the de-excitation of the beam particles are collected with a measureable Doppler shifted wavelength λ given by

$$\lambda = \lambda_0 \frac{1-\beta \cos\alpha}{\sqrt{1-\beta^2}}, \text{ where } \beta = \frac{v}{c} = \sqrt{1 - \left(\frac{mc^2}{mc^2 + eU_{tot}}\right)^2} \qquad (1)$$

where $\lambda_0$ is the unshifted $H_\alpha/D_\alpha$ wavelength and α is the angle between the beam particles' trajectory and the direction of the collected photon, as schematized in fig. 1. β is the ratio of the beam particles speed v and the speed of light c: it is calculated from the beam particle mass m, the speed of light, the electron/$H^-$ charge e and from $U_{tot}$, which is the total potential difference in the acceleration system.

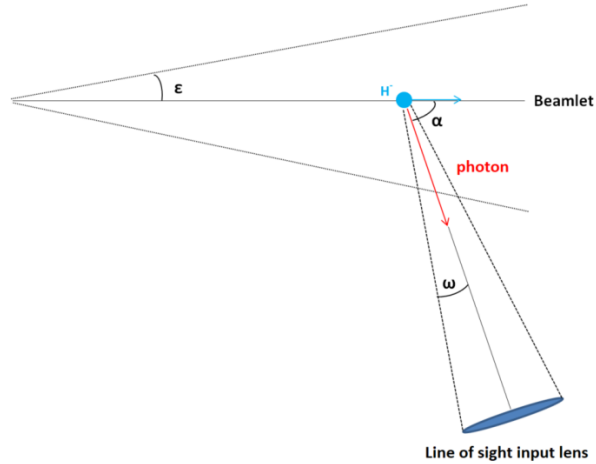

FIG. 1. Schematic representation of the angular quantities α, ε and ω.

In the BES diagnostic, the light collected by the collimators is fed by means of optical fibers to a spectrometer; an example of BES spectrum collected in SPIDER is shown in fig. 2. The unshifted $H_\alpha$ emission is visible together with the full energy Doppler peak at about 655.25 nm, which represents the emission of most beam particles at full energy, in this example 21.5 keV. BES spectra also show Doppler emissions related to lower energies compared to the total energy. They are given by the negative ions which were neutralized inside the acceleration system (the so called stripping losses). While these emissions are barely detectable in the case of fig. 2, this is generally not true.

To measure the beam-line of sight direction α, Gaussian fits (red curves in fig. 2) are applied to the full energy Doppler peak and the unshifted $H_\alpha$ peak to measure λ and $λ_0$; while $λ_0$ is by itself known by quantum physics, the measured value helps correcting potential offsets in the wavelength calibration of the spectrometer. Due to the presence of stripping losses emissions in the spectrum, the wavelength intervals dedicated to the fits of the full energy Doppler peak and of the unshifted peak are limited to the minimum range which can provide stable results in most of BES spectra. The extensions of the red curves in fig. 2 indicate the effective wavelength intervals that were selected for the Gaussian fits in this case. The beam direction can be measured in terms of α inverting eq. 2, with λ given by the centroid of the Doppler peak Gaussian fit.

$$\alpha = \cos^{-1}\left(\frac{\lambda_0 - \lambda\sqrt{1-\beta^2}}{\beta \lambda_0}\right) \qquad (2)$$

In SPIDER, the acceleration system is composed by the Plasma Grid (PG) facing the source, the Extraction Grid (EG) and the final Grounded Grid (GG). To calculate β, the total potential energy $U_{tot}$ is obtained from the sum of the potential differences in the PG-EG extraction gap (i.e. extraction voltage $U_{ex}$) and in the EG-GG acceleration gap (i.e. acceleration voltage $U_{acc}$), which are separately controlled by dedicated power supplies. $U_{tot}$ is averaged over the time interval of the spectrum acquisition, with proper correction for background offsets in the readings of the power supplies.

The Gaussian curve fit of the full energy Doppler peak also gives the integral of the peak, which can be compared from LoS to LoS to get a spatial profile of the beam intensity.

From the full energy Doppler peak it is also possible to get the beam divergence ε; the fact that the observed beam particles have a finite angular distribution involves that also α of the collected photons has an equivalent distribution. This in turn implies a broadening of the full energy Doppler peak, that once measured allows to estimate the beam divergence ε. The width W of the full energy Doppler peak is available from the applied Gaussian fit. From this width, besides the $H_\alpha/D_\alpha$ intrinsic linewidth $\Delta\lambda_N$, several broadening factors must be subtracted: the broadening introduced by the spectrometer instrumental function $\Delta\lambda_I$, the collection angular spread ω introduced by the finite dimensions of the collimator optical aperture (as schematized in fig. 1) and the ripple of the total acceleration voltage υ. Following the model described in ref. 11, the e-folding beam divergence ε can be calculated as follows:

$$\varepsilon = \sqrt{2}\left(\frac{\lambda_0 \beta \sin\alpha}{\sqrt{1-\beta^2}}\right)^{-1} \sqrt{W^2 - \Delta\lambda_N^2 - \Delta\lambda_I^2 - \left(\frac{\lambda_0 \beta \sin\alpha}{\sqrt{1-\beta^2}}\right)^2 \omega^2 - \left[\frac{e\lambda_0^2}{mc^2\beta}(\beta - \cos\alpha)\right]v^2}. \qquad (3)$$

W and the broadening factor under the square root of eq. 3 are meant to be estimated in terms of sigma-RMS widths for a Gaussian curve. The value of α in eq. 3 is measured following eq. 2. υ is estimated from the standard deviations of the voltage signals of the power supplies feeding the grids in the acceleration system; the standard deviations are computed in the time interval of the spectrum acquisition. Too large values of $\upsilon$ (above 10 % of V) lead to discarding the acquired spectra, since they imply the presence of drifts in total acceleration voltage or of voltage breakdowns.

In the example BES spectrum of fig. 2, W=80.9 pm, the four contributions to be quadratically subtracted amount to 13.2 pm (instrinsic linewidth), 19.0 pm (instrumental function width), 6.6 pm (broadening caused by collection angle) and 5.7 pm (broadening caused by voltage fluctuations), respectively. The calculation of the measurement errors is performed according to the model presented by B. Zaniol in ref. 11 . Table 1 lists the absolute values and the precision of the parameters requested to calculate $\alpha$, $\varepsilon$ and their measurement errors. In the case of the BES spectrum in fig. 2, the resulting measurements of beam direction and divergence are $\alpha=(74.6\pm0.1)°$, while $\varepsilon=(24.6\pm0.2)$ mrad e-folding, respectively.

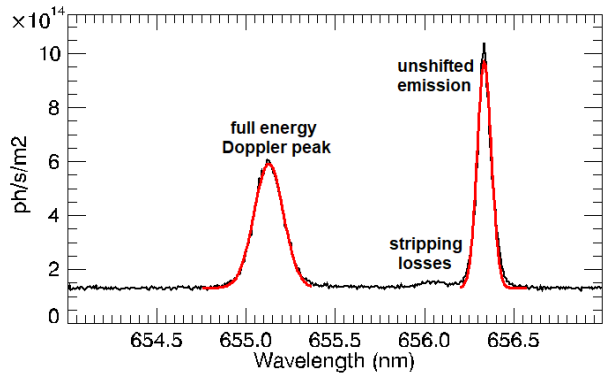

FIG. 2. Typical example of BES spectrum acquired in SPIDER (4.7 s photon collection time). The black curve indicates the acquired spectrum, while the red lines indicate the Gaussian curves fitted on the peaks. Source input power was 320 kW, source gas pressure was 0.4 Pa, $U_{ex}$=1.5 kV, $U_{acc}$=20 kV; extracted beam current density was 14 A/m². According to the presented analysis method, $\alpha=(74.6\pm0.1)°$, while $\varepsilon=(24.6\pm0.2)$ mrad e-folding.

| Quantity | Value |
| --- | --- |
| $\lambda_0$ (unshifted $H_\alpha/D_\alpha$ wavelength) | 656.297 nm/656.103 nm |
| Typical error in the estimation of $\lambda$ or $\lambda_0$ in the acquired spectrum | 0.7 pm÷4.0 pm |
| Rest mass of $H^-/D^-$ ions ($mc^2$) | $9.388\cdot10^8$ eV / $1.876\cdot10^9$ eV |
| Relative error on V measurements | 1 % |
| Typical error on W measurements | 0.7 pm÷5 pm |
| $\Delta\lambda_N$ (intrinsic linewidth) | 13.2 pm |
| Typical $\Delta\lambda_I$ (depending on the CCD region) | 19 pm÷27 pm |
| Typical error on $\Delta\lambda_I$ (depending on the line of sight) | 0.3 pm÷1.0 pm |
| ω (observation angle) | 1.5 mrad |
| Relative error on ω, due to its variation along the lines of sight | 5 % |
| Typical value of υ (ripple in total acceleration voltage) | 1÷3 % |
| Relative error in the estimation of υ | 10 % |

Table 1: Values used in the estimation of α, ε and their measurement errors. Typical values refer to quantities that are measured case by case or vary for each line of sight; otherwise, values are meant to be constants in the calculation. Values referred to peak widths or spectral broadening factors are meant to be estimated as sigma-RMS widths for a Gaussian curve.

### III. THE BES DIAGNOSTIC IN SPIDER

#### A. The diagnostic hardware

The light emitted from the interaction between beam and gas is collected by several optic heads, composed by metal cylinders, each one hosting a plano-convex lens with 10 mm clear aperture and 120 mm focal length. The light is focused on the head of a high OH content silica-silica fiber, with 400 μm core diameter and 0.22 numerical aperture. In each optic head the FC/PC fiber connector is fastened to a fiber holder, which was aligned in laboratory so that the conjugate plane of the fiber head is at 3 m distance from the lens, that is beyond the distance of the observed beam[10,26]. The optical heads are mounted on dedicated supports, which allow to properly align them with horizontal/vertical translations and rotations on three axes. Up to three optical heads can be installed in each support.

The light collected from each optic head is carried by a sequence of three optical fibers patchcords ( 38 m or 44 m long), from the SPIDER vessel to the outside of its concrete bioshield, and then to a separate building, where spectrometers are installed. To analyze the spectrum of the collected light, a Princeton Instruments Isoplane SCT-320 spectrometer is employed. This spectrometer is characterized by an innovative design, with higher throughput and lower off-axis aberrations than a conventional Czerny-Turner configuration. The spectrometer mounts three diffraction gratings, with 150 gr./mm (for survey purposes), 2400 gr./mm (for BES operation) and 3600 gr./mm. These values of groove spacing were selected not just according to the BES diagnostic requirements, but also compatibly to the needs of the Optical Emission Spectroscopy diagnostic, which analyzes the light emitted from the source plasma[27]. While 150 gr./mm gratings allow a low resolution spectral survey of the entire visible range, the 3600 gr./mm grating allows to better resolve molecular emission spectra. The 2400gr./mm grating is used by default for BES; all the data presented in this paper were obtained using this grating. The spectrometer is equipped with Princeton Instruments back-illuminated CCDs PIXIS 1024B, each one with 1024x1024 pixels, 13 μm wide. At the entrance slit of the spectrometer, which is 50 μm wide, up to 22 fiber ends are vertically stacked so that the corresponding spectra can be simultaneously acquired by the 2D CCD at the exit of the spectrometer. The spectrometer spectral dispersion is 7.6 pm/pixel, compatibly with the original design of the diagnostic[10]; the instrumental function width ranges from 19 pm to 27 pm(sigma-RMS). All the lines of sight are calibrated in wavelength. A first calibration was performed with a Hg-Cd lamp; a finer calibration was performed around the $H_\alpha$ wavelength with a hydrogen lamp, a neon lamp and a cadmium lamp, to provide calibration lines at 650.65 nm (Ne I), 652.21 nm (Cd I 2$^{nd}$ order), 653.29 (Ne II) and 656.28 nm ($H_\alpha$ itself). Each optical line from the optic head to the spectrometer (i.e. only excluding the silica vacuum vessel windows) was calibrated in intensity, except for the vertical LoSs because of technical difficulties. The calibration was performed using the optical setup to acquire the light emitted by an Ulbricht sphere, and comparing the measured spectrum with the known spectral emission curve of the sphere.

Control and data acquisition of BES cameras and spectrometers are managed by the SPIDER MDSplus software platform[28], which is also used to store the BES analysis results.

### B. The geometry of the lines of sight

Each support dedicated to the BES optic heads is installed in front of one of the several silica BES viewports, placed downstream the exit of the acceleration system and allowing to look at the beam with two different angles. Four viewports on the north lateral side are placed at about 90 cm distance along the machine axis from the final grid of the acceleration system, the Grounded Grid (GG). As many viewports are present on the south side of the vessel at the same distance from the GG, at interlaced vertical positions with respect to the ones in the opposite side. The optical heads mounted in front of these windows are oriented so that α=75°, i.e. slightly counter-beam. At a similar distance from the GG along the machine axis, four viewports are placed on the top of the vessel, again with α=75° but on the vertical plane. In total, in proximity of the acceleration system a maximum number of 24 horizontal LoSs and 12 vertical LoSs is available. A second set of 8 viewports is also present on the front lid of SPIDER vacuum vessel, providing maximum 24 horizontal Lines of Sight (LoSs) at α=45° from the sides of the beam dump. Temporarily no optic heads are installed in front of these viewports, since their LoSs would be blocked from the CFC tiles of the instrumental calorimeter STRIKE[8]. Schematic top and side views of the SPIDER main components, together with the aiming of horizontal and vertical lines of sight (LoS), are shown in fig. 3. Each group of LoSs is represented as a single one, being the LoSs of each group vertically or horizontally stacked. The LoSs are labelled as NB (north – beam), SB (south – beam) or TB (top – beam), followed by a progressive number increasing from top to bottom or left to right (defined looking downstream).

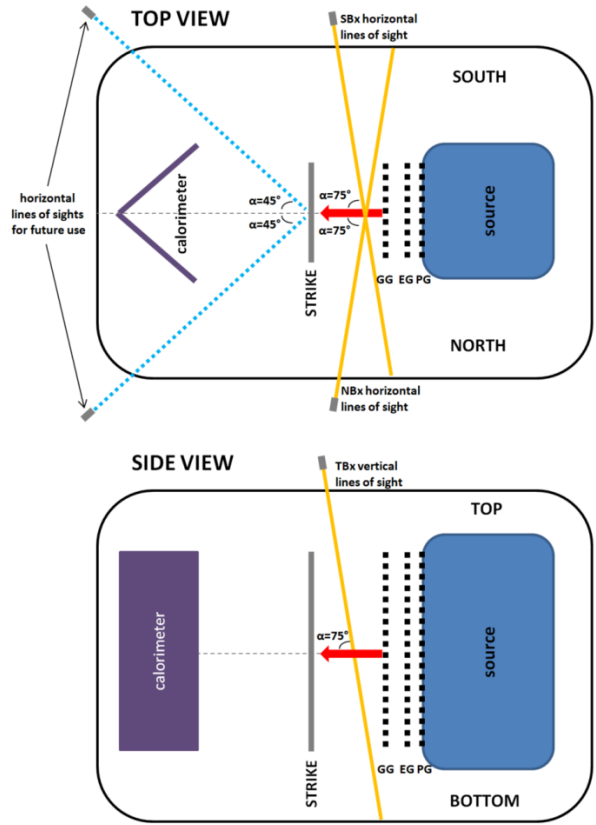

FIG. 3. Schematic side and top representation of the main SPIDER components and of the vertical and horizontal BES lines of sight. Each line of sight in the figure actually indicates a group of lines of sight, horizontally or vertically stacked.

The data collected from the several LoSs must be interpreted taking into account the layout of the LoSs themselves with respect to the beam position. The grids composing the acceleration system have 1280 apertures, through which as many beamlets are generated. Apertures are spatially divided in 16 groups of 16 (vertical) x 5 (horizontal); this is shown in fig. 4, which presents a scheme of the PG apertures as seen from the source side. The spacing between apertures inside each group is 20 mm horizontally and 22 mm vertically. The BES optic heads supports were designed to keep the double of these spacings between vertical and horizontal optic heads, respectively. In other words, horizontal/vertical optic heads mounted on the same support can be aligned to observe either even or odd rows/columns of beamlets. The optic heads were manually aligned in the nominal plane of the rows/columns of apertures axes, using reference points on the viewports, on the vessel or on the accelerator surfaces extracted by the 3D CAD model of SPIDER.

The positioning and identification label of the active lines of sight is shown in fig. 4; LoSs are represented as light blue bars. The choice was driven by the decision of closing most of the grids apertures with a Mo mask to limit the gas flux from the source[4]. The 80 still active apertures are indicated in fig. 4 with red circles. All the vertical (TBx) lines of sight intercept the columns of beamlets in line. The horizontal LoSs with odd numbering (eg. NB1 or SB3) intercept rows of beamlets in line, while LoSs with even numbering (eg. NB2 or SB2) can only observe the external sides of adjacent beamlets.

By measuring α from the horizontal BES lines it is possible to study the horizontal deflections that are caused by the alternated vertical magnetic fields generated by permanent magnets embedded in the EG; the purpose of these magnetic fields is to deviate the electrons co-extracted from the source and dump them on the EG[25]. The resulting deflections are in the same direction within each row of apertures and opposite between adjacent rows. This pattern of deflections is indicated in fig. 4 for the active beamlets, using blue arrows (leftward looking from the source, or equivalently in north direction) or red arrows (rightward – in south direction). It is worth noting that with this arrangement all the NBx LoSs in line with open apertures (NB1, NB4, NB7, NB10) observe beamlets that deviate leftward, while the SBx LoSs in line with open apertures (SB3, SB6, SB9, SB12) observe beamlets that deviate rightward. As consequence, in all the just mentioned LoSs the observed beamlets are deflected towards the optic heads, therefore the beam-LoS angle α is expected to be lower than 75 °. The opposite happens for the horizontal LoSs (NB2, NB5, NB8, NB11, SB2, SB5, SB8 and SB11) intercepting the active beamlets only sideways; in this case the expected deflections cause an increase of α.

In order to correct the horizontal beamlet deflections, two (plus one) compensation methods were implemented in the segments (Fig. 4) that compose the SPIDER grids[25,29]. In segment 1, 3 and 4 the compensation is provided by magnets embedded in the GG, although the performances of this system are spoiled by the fact that the magnets in the EG were all erroneously mounted by the manufacturer in reversed direction. In the bottom half of segment 2 the compensation is provided by an offset in the aperture positions at the GG (electrical compensation); at last, in the top half of segment 2 no compensation was implemented at all to provide a reference case. A first comparison of these compensation methods by means of BES results is presented in the following section.

Horizontal deviations can be studied in detail thanks to the limited number and the distance between active beamlets, so that no overlapping beamlets from adjacent rows may be seen by a line of sight. Otherwise, for deflection values of some mrads, the full energy Doppler peaks corresponding to those beamlets would not be distinguishable, and a single broader peak in BES spectra would be measured, with a consequent overestimation of beam divergence.

In addition to the horizontal alternated deflections, all the beamlets can be deflected in vertical direction, because of a drift of plasma particles in the proximity of the PG. This drift is caused by a horizontal magnetic filter field, produced by a current flowing in the PG; the purpose of this magnetic field is to reduce density and speed of the electrons in proximity of the PG[1,30], thus reducing the amount of co-extracted electrons. This phenomenon is mostly studied from the results of vertical BES lines of sight.

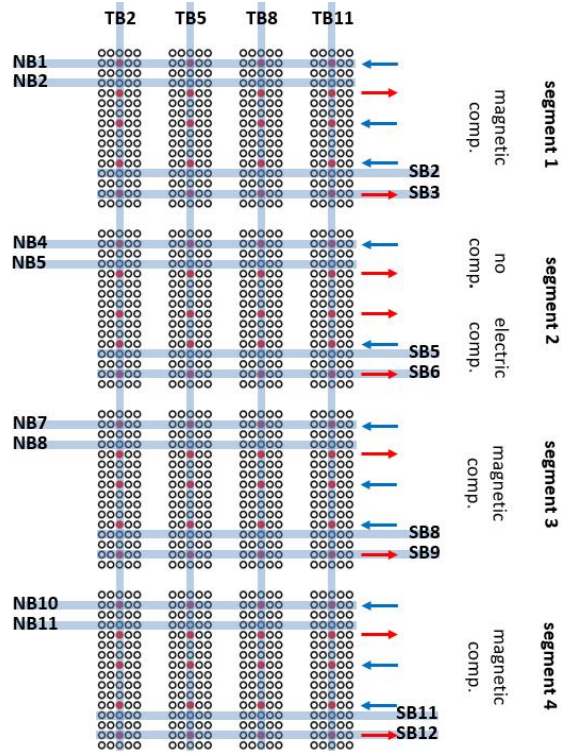

FIG. 4. Schematic representation of the Plasma Grid apertures, looking from the source. Red spots indicate the grid apertures left open by the Mo mask; BES LOSs are indicated in light blue. Blue and red arrows indicate the direction of the deflections on the active beamlets, caused by the magnetic field generated by the EG magnets; blue arrows indicate left/north deflections, red arrows indicate right/south deflections.

## IV. FIRST RESULTS

The results of the BES diagnostic here presented belong to the first campaign with beam extraction, performed in 2019 with hydrogen and without Cs evaporation in the source. The most relevant data, that are here reported and discussed, were obtained by varying the extraction voltage between 0.75 kV and 3.5 kV, at acceleration voltage values of 20 kV, 25 kV and 30 kV. The current $I_{PG}$, flowing in the PG grid to produce a magnetic filter field which reduces the amount of co-extracted electrons[1,30], was set at 1.5 kA or 3.0 kA, corresponding to 2.4 mT and 4.8 mT in proximity of the PG. A scan of $I_{PG}$ is also considered, from 1.0 kA to 3.5 kA in steps of 0.5 A, with 1.5 kV extraction voltage and 20 kV acceleration voltage. In all the considered SPIDER beam pulses a total radiofrequency power of 320 kW was used to generate the plasma in the source. Given the limited number of active beamlets and the low extracted beam current density (in the order of 10 A/m$^2$), to get spectra with sufficient signal-to-noise ratio the exposure time of the spectrometer CCD was set at 4.7 s and acquisitions were repeated every 5 s.

The analysis of BES spectra was performed as described in sec. II. From a qualitative point of view, differently from what observed by others[23][24], in the SPIDER experimental conditions considered in this paper the full energy Doppler peak does not present large wings at the base, with respect to a Gaussian shape. While in the SPIDER BES spectra of vertical LoSs the full energy Doppler peak can be almost always considered Gaussian, in some cases the spectra of horizontal LoSs showed slightly asymmetrical shapes. This effect is related to a modification of the beam particles velocity distribution in the horizontal direction; this distribution was in fact affected by the magnetic fields that are present in the acceleration system to block co-extracted electrons or to compensate the beamlets aiming[8,25]. This deformation of beamlets was also observed by IR imaging of the instrumented SPIDER calorimeter[8]. The phenomenon was generally observed in conditions of not optimized beam divergence, however it was not always simultaneously observed in all horizontal LoSs.

The quantitative results obtained for each parameter estimated via BES are here separately presented and discussed.

**A. Beam direction measurements**

It is known that the magnetic filter field, generated by the PG current to reduce the amount of co-extracted electrons, causes an overall deflection of the negative ion beam[31-33]. Varying $I_{PG}$ from 1.0 kA to 3.5 kA, the values of α measured from the TB5 (vertical) LoS showed a downwards deflection of (7±2) mrad. This is shown in fig. 5, in which beamlet deflection, calculated as the difference of TB5 α measurements with the nominal value, are plotted as a function of $I_{PG}$. The error bars in the plot indicate the statistical errors of the measurements, which are less than ±2 mrad, while the orange bar represents the range of the offset that may have been introduced by the positioning of the supports of optical heads, which is max. ±9 mrad. From the collected data it can be concluded that the beam vertical deflection is acceptably low in the described conditions.

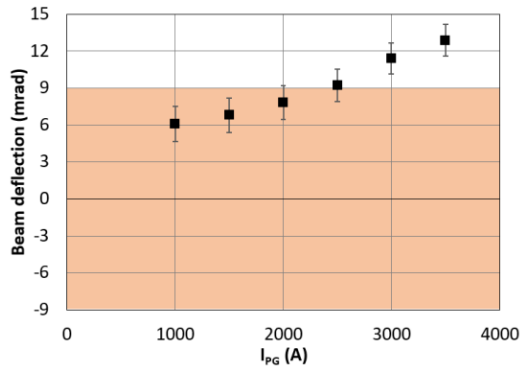

FIG. 5. Beam deviation measurements from the vertical LoSs TB5 as a function of the plasma grid current $I_{PG}$. The orange bar indicates the maximum interval of systematic deflections that may have been introduced by the manual alignment of optic heads supports.

From BES measurements it is also possible to characterize the beamlet horizontal deflection, as a function of extraction voltage, acceleration voltage and deflection compensation mode. Fig. 6 shows the beamlet deflection (full symbols) measured from the horizontal LoS NB7, the LoS receiving the highest light intensity on the northern side from beamlets of segment 3 (magnetic compensation), as function of extraction voltage, at 20 kV (plot a), 25 kV (plot b) and 30 kV (plot c) acceleration voltage. Analogously, plots d, e and f of fig. 6 show the values of beamlets deflection (full symbols) measured from LoS SB9, that is the LoS receiving the highest light intensity on the southern side from the beamlets of that same segment. As in fig. 5, the range of the offsets that may affect the measurements of each LoS is indicated in the plots of fig. 6 with orange bars. What emerges from experimental results is firstly that increasing the extraction voltage leads to a compensation or even overcompensation, i.e. inverting the north/south deflections, of the beamlets aiming. In the case of NB7, the variation of beamlets deflection is up to 29 mrad; this phenomenon appears weaker for SB9 measurements, but it is still measurable. Data from NB7 also show the deflection decreases by increasing acceleration voltage, by about 7÷10 mrad between 30 kV and 20 kV; the effect is however not distinguishable in SB9 data.

From the first collected data it is possible to make a rough comparison between the directions of beamlets resulting from different techniques of deflection compensation. From the top half of segment 2, with no deflection compensation, the intensity of the NB5 LoS was sufficient to provide detectable spectra, while for the bottom half of segment 2, with electric compensation, detectable spectra were collected by the SB6 LoS. Deflection values of NB5 and SB6 are shown in fig. 6 (plots a,b,c and d,e,f, respectively) with empty symbols as functions of extraction voltage, for the same plasma pulses as the other shown data. As explained in sec. III, the trend of α and then of deflection values for NB5 should be opposite with respect to SB6; this is interestingly confirmed only at low values of extraction voltage. The offset between the data with electric or no compensation and those with magnetic compensation is compatible with the alignment precision of optic heads supports, so potential differences in effects of aiming compensation may be the result of offsets within ±9 mrad.

A sufficiently accurate measurement of the relative deflection between beamlets aiming in opposite directions was possible by comparing the data of LoSs SB2 and SB3, looking at the bottom of segment 1 (magnetic compensation); their optic heads are indeed mounted on the same mechanical support, so they should have a lower misalignment between themselves. Together with NB5, SB2 was the only LoS not directly intercepting beamlets that was able to provide spectra with sufficient Doppler emission intensity. As example, Fig. 7 shows the difference between the values of α measured by SB3 and SB2, as a function of extraction voltage, at 20 kV acceleration voltage and two values of magnetic filter field current (1.5 kA and 3.0 kA). As expected, the difference of α values is negative: beamlets observed by SB3 are deflected in southern direction (lowering α), vice versa for SB2 (increasing α); the absolute values tend to reduce linearly from about 50÷60 mrad towards an alignment with the machine axis. These results are essentially in agreement with what measured by SPIDER diagnostic calorimeter STRIKE and reported in ref. 8 .

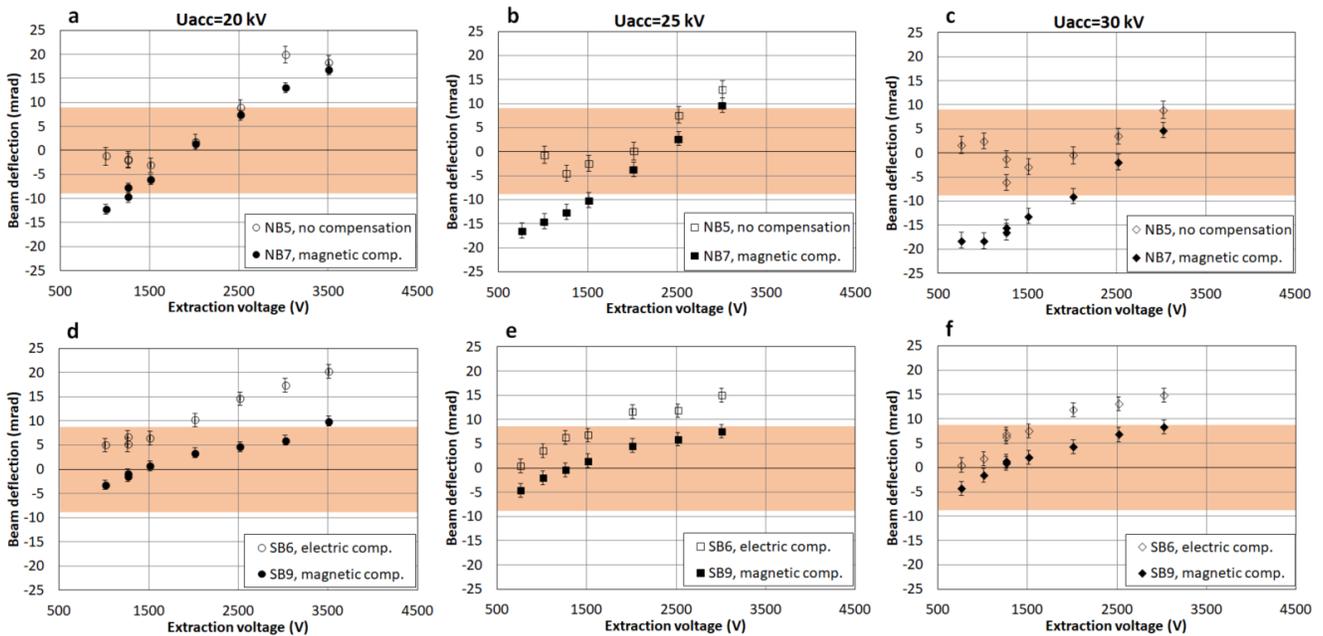

FIG. 6. Beam deviation measurements from LoSs NB5 (segment 2 – no deflection compensation), NB7 (segment 3 – magnetic compensation), SB6 (segment 3 – magnetic compensation) and SB9 (segment 3 – magnetic compensation), as function of extraction voltage. NB5 and NB7 data are shown in plots a, b and c, while SB6 and SB9 data are shown in plots d, e and f. Data were collected at three values of $U_{acc}$ voltage: 20 kV (plots a and d), 25 kV (plots b and e) and 30 kV(plots c and f); for all measurements, $I_{PG}$ was 3.0 kA. Orange bars indicate the maximum interval of systematic deflections that may have been introduced by the manual alignment of optic heads supports.

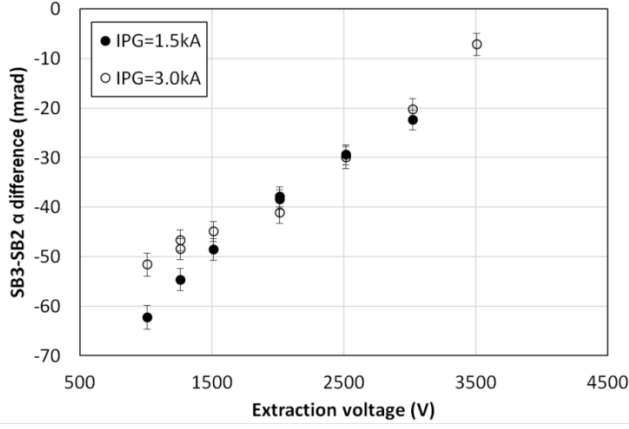

FIG. 7. Difference between the values of α in LoS SB3 and SB2, as function of extraction voltage. Data were collected at 20 kV $U_{acc}$ voltage and for two values of $I_{PG}$: 1.5 kA (full symbols) and 3.0 kA (empty symbols).

## B. Beam divergence measurements

The beam divergence is known to be mainly dependent on the perveance P of the extraction gap[22], which is here estimated as

$$P = \frac{I_H + I_e\sqrt{m_e/m_H}}{U_{ex}^{3/2}}, \quad (4)$$

where $I_H$ is the extracted beam current, $I_e$ is the co-extracted electron current, $m_e$ and $m_H$ are the electron and ion mass, and $U_{ex}$ is the extraction voltage. P is then normalized to the space charge limited level $P_0$, as given by the Child-Langmuir law. To estimate P, $I_H$ was assumed to be the current of the power supply for the acceleration gap (EG-GG), corrected by a factor 0.4, as calculated for SPIDER by comparing the beam electric and calorimetric measurements from the instrumented calorimeter[8], in order to exclude stripping and secondary electrons from the estimation. $I_e$ and $U_{ex}$ are estimated from the power supply for the PG-EG gap. Plots a and b of fig. 8 show the dependency of beam divergence (in e-folding mrad) on the normalized perveance $P/P_0$, at $I_{PG}$=1.5 kA and $I_{PG}$=3.0 kA respectively, with 20 kV acceleration voltage. $P/P_0$ was varied by changing $U_{ex}$ from 0.75 kV to 2.5 kV ($I_{PG}$=1.5 kA) and from 1.5 kV to 3.5 kV ($I_{PG}$=3.0 kA). The plots show the horizontal beam divergence, estimated as average of SB3, SB9 and SB12 (magnetic deflection) measurements, and vertical divergence, as measured from TB5. The error bars of divergence are not visible because they are smaller than the data markers. The first important result, confirmed by beam calorimetry[8], is that beamlets are horizontally elongated, as the horizontal divergence is larger than the vertical one. The minimum achievable vertical beam divergence was about 20 mrad and 30 mrad, for the two $I_{PG}$ values; the divergence values are in agreement with the measurements of the instrumented calorimeter and with the visible imaging diagnostic[8,9]. As explained in sec. II, the difference between horizontal and vertical beam divergence, which exceeds 12 mrad, has to be entirely attributed to aberrations of beam optics, caused by the EG magnetic field and accentuated by the fact that the acceleration system is operated at voltages that are 1/5-1/3 of the target one (100 kV). Beam divergence is minimized at $P/P_0 \approx 0.25$ at $I_{PG}$=1.5 kA and about 0.18 at $I_{PG}$=3.0 kA. This difference may be due to the density of negative ions available in front of the PG apertures, and then to the meniscus shape; increasing the magnetic filter field current also reduces H- density[31-33] and then the extracted current (see next section) in SPIDER.

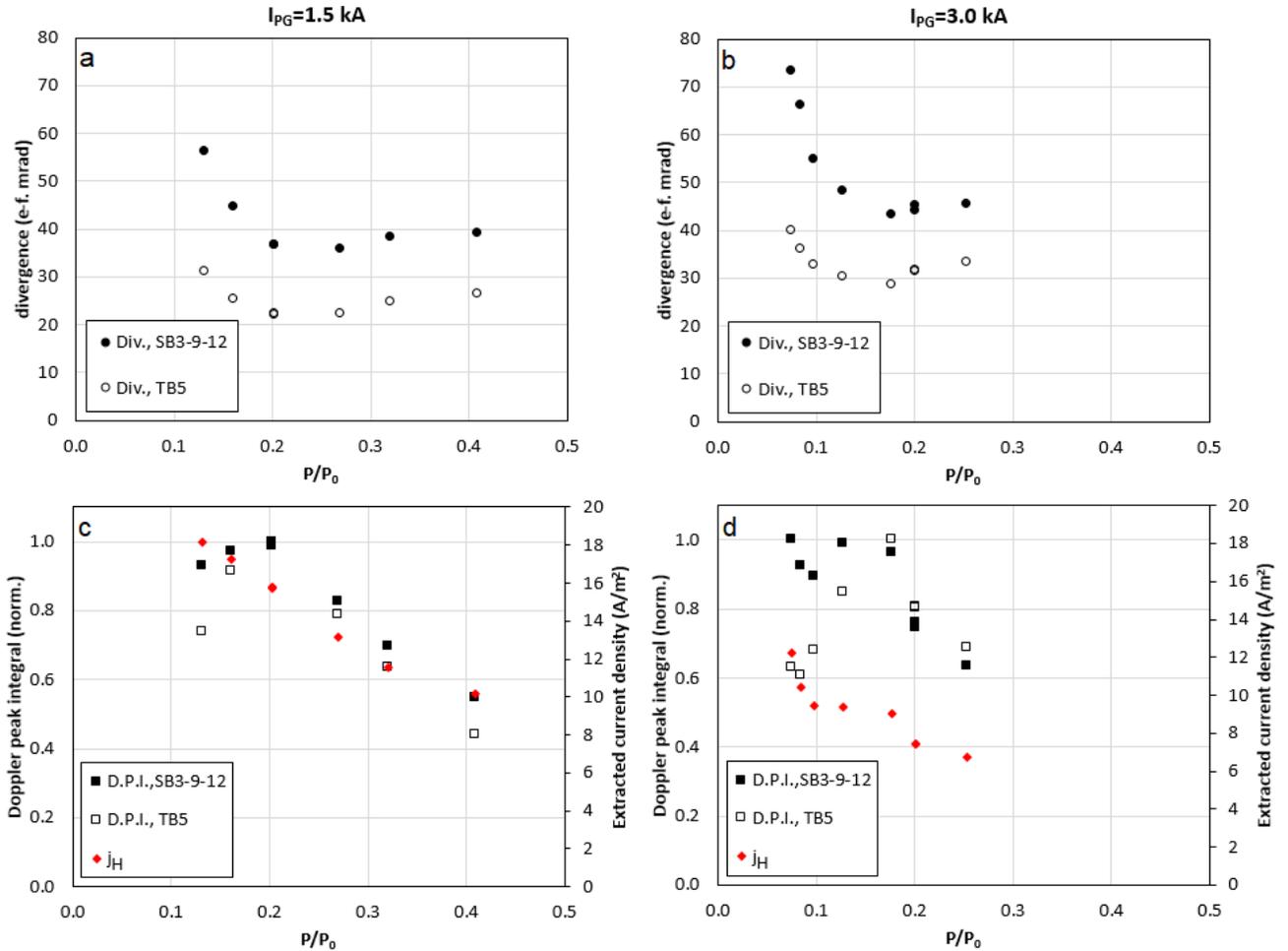

FIG. 8. Plot a: Horizontal beam divergence (calculated as average of SB3, SB9 and SB12 measurements) and vertical beam divergence (as measured from TB5) as function of normalized perveance, during an extraction voltage scan with $U_{acc}$=20 kV and $I_{PG}$=1.5 kA. Plot b: Horizontal and vertical beam divergence (estimated as in plot a) as function of normalized perveance, during an extraction voltage scan with $U_{acc}$=20 kV and $I_{PG}$=3.0 kA. Plot c and d: Full energy Doppler peak integral, measured as average from LoS SB3, SB9 and SB12, and from TB5, as function of normalized perveance; the extracted current density is also plotted. The data refer to the same cases examined in plot a and b, respectively.

## C. Beam intensity measurements

Besides beam direction and divergence, the intensity of the full energy Doppler peak in BES spectra can provide information about beam intensity and uniformity. In the present BES experimental setup the alignment between LoS and beamlets proved to be too critical to allow comparison of Doppler peak intensity between different LoSs. A spatial distribution of the beam is not possible yet in this setup. At the average distance from the GG at which beamlets and LoS intersect (about 0.35 m), the LoS extension, around 10 mm, is comparable to the beamlet dimensions, in the order of few tens of millimeters; beamlets deflections can modify the distance between the axes of beamlets and LoS by about 0.34 mm per milliradian of deflection. This issue might get self-solved when all the 1280 PG apertures will be active, so that the mixing between different rows and columns of beamlets will provide a more uniform beam spatial distribution to sample. Alternatively, remotely controlled systems to move the BES optic heads supports may be tested. This issue is in any case expected to be much less severe on the farther $\alpha=45°$ LoSs, which will become available when the STRIKE calorimeter will be moved to the rest position; at the position of those LoSs, indeed, the beamlet dimensions should be larger, therefore the full energy Doppler peak intensity data will be less sensitive to the distance between the axes of beamlets and LoSs.

The full energy Doppler peak integral can still provide useful information about the beam current intensity variations due to changes in the source or acceleration system parameters. As example, plots c and d of fig. 8 show the full energy Doppler peak integral and the extracted current density $j_H$ ($I_H$ divided by the grid apertures area), for the same experimental cases shown in the respective above plots a and b, that is for the two conditions with $I_{PG}=1.5$ kA and $I_{PG}=3.0$ kA. Similarly to plots a and b, the shown Doppler peak integral estimates are measured from TB5 and as average between SB3, SB9 and SB12 results; values are in arbitrary units because it was not possible to perform the intensity calibration of TBx LoSs. Error bars are not visible in the plot because they are smaller than the data markers. The full energy Doppler peak integral linearly depends on the extracted current density, but also on the beamlets position with respect to the line of sight and on beamlets divergence: the larger the beamlet, the lower the fraction of its volume that is intercepted by the line of sight. Depending on the LoS alignment, the full energy Doppler peak intensity on analyzable spectra ranged from $5.3 \cdot 10^{13}$ to $7.3 \cdot 10^{14}$ photons/m²/s per 1 A/m² of extracted current density. The full energy Doppler peak intensity would also depend on the excitation cross section of beam particles and in turn on beam energy, but no relevant cross section variations are expected in this limited $U_{ex}$ variation range[35,36].

As shown in plots c and d of fig. 8, the Doppler peak integral measurements generally well follow the trend of the extracted current density. However, the agreement between the trend of $j_H$ and Doppler peak integral is not verified at low values of $P/P_0$, especially for TB5. This can be explained, as mentioned above, with the relationship between beamlets divergence and collected light intensity. It is worth noting that the agreement worsens in both $I_{PG}$ cases when beam divergence is above 45 mrad. It cannot be excluded, however, that the changes in horizontal beamlet deflections induced by the changes in $U_{ex}$ play a significant role in the alignment between beamlets and vertical LoSs.

Another interesting fact is that $j_H$ decreases with $P/P_0$, despite P linearly depends on $I_H$ and then on $j_H$ (extracted current increases with $U_{ex}$, but the inverse proportionality of $U_{ex}^{3/2}$ on P estimation is dominant). Furthermore, when $I_{PG}$=3.0 kA, $j_H$ is not only lower, but an increase is observed at the $P/P_0$ value which minimizes beam divergence. This effect is only barely visible in the $I_{PG}$=3.0 kA case. This may indicate a partial blocking of the beamlets at the GG level at non optimal values of $P/P_0$, because of excessive beamlets deflections and/or too large beam divergence.

## V. CONCLUSIONS

In this paper, the final implementation of the BES diagnostic in SPIDER is presented. The diagnostic LoSs distribution were correlated to the position and the physical properties of the H$^-$ beamlets available from the acceleration system. The reduction in SPIDER from 1280 to 80 available beamlets was necessary to reduce gas conductance from the source, so to keep gas pressure around the source low enough for proper source operation[4]. This choice has clearly affected not only the signal-to-noise ratio of BES spectra but, above all, the effective number of LoSs collecting light; it also amplified the effects of beamlets-LoS alignment on the measured full energy Doppler peak. Nevertheless, the first results of BES data, here presented and discussed, allowed a preliminary characterization of the three different techniques of beamlet deflection compensation installed in SPIDER (including the no compensation reference case). By increasing the extraction voltage, the reduction of beamlet deflection is evident and comparable among the three methods. Increasing the acceleration voltage between 20 kV and 30 kV did not lead to a visible improvement of deflection compensations. In agreement with the observations of the STRIKE IR imaging diagnostic,[8] the deflection values given by the magnetic and electric compensation methods did not differ from the no compensation reference case by more than 15 mrad.

Reducing the opposing deflections of the rows of beamlets is crucial to limit the overall beam divergence. A comparison of BES measurements with a single beamlets and with multiple beamlets with opposing deflections has been recently performed in the BUG test facility of IPP Garching. The experimental results, showed that the overall beam divergence, as measured by BES, can increase up to a factor of 3 because of the so called criss cross deflections. In the case of SPIDER, the shown data demonstrate that the difference between opposite deflections can be reduced below 10 mrad even in conditions that are far from those upon which the acceleration system was designed.

It was also possible to characterize how beam divergence depends on the normalized perveance, in a regime of H$^-$ Cs-free production and under different operation conditions. Absolute values of beam divergence reached minimum values around 20 mrad e-folding; compared to the SPIDER target of 7 mrad this is a promising result, considering the unoptimized conditions of the acceleration system. The range of measured beam divergence values is comparable to the measurements obtained in other RF negative ion sources[14,15,23], with cesium evaporation (that leads to less co-extracted electrons and one order of magnitude higher extracted current density) and beam optics closer to the design targets of the respective acceleration systems.

The measurements of Doppler peak integral resulted to be in good proportionality with the extracted current density, except in cases of too large beam divergence. The performances of the present BES diagnostic were not sufficient to provide reliable profiles of full energy Doppler peak intensity, from which information about beam uniformity could be derived. This may be solved in future by adding remotely controlled motorized translators to the BES optic heads. Also, an accurate cross check between BES information and STRIKE calorimetric data of each beamlet will be performed. This check will be made possible by means of the dBES code[26,37-38], which will simulate BES spectra from the information provided by the diagnostic calorimeter STRIKE; the simulations may also help in deriving estimates of beam current density from the measurements of Doppler peak integral. It should also be highlighted that when beam power or beam duration will significantly increase, the panels of the diagnostic calorimeter will be opened and moved to the safety position, to let the beam hit the main SPIDER water cooled calorimeter. In this condition, not only the present LoSs but also the front lid LoSs, farther and then more robust in terms of beamlet-LoS alignment (the higher the distance the higher the beamlets transversal dimensions), will be available. BES will then play a major role in characterizing the properties of the beam when SPIDER will get closer to its final performances: 100 keV, 350 A/m$^2$ (H)/285 A/m$^2$ (D), from 1280 available beamlets.

## AVAILABILITY OF DATA

The data that support the findings of this study are available from the corresponding author upon reasonable request.

## ACKNOWLEDGMENTS

The work leading to this publication has been funded partially by Fusion for Energy. This publication reflects the views only of the authors, and F4E cannot be held responsible for any use which may be made of the information contained therein. The views and opinions expressed herein do not necessarily reflect those of the ITER Organization.